\documentclass{pnastwo}

\pdfoutput=1

\usepackage{float}
\usepackage{epsfig}
\usepackage{lscape}
\usepackage{enumerate}
\usepackage{multirow}
\usepackage{array}
\usepackage{comment}

\usepackage{amssymb,amsfonts,amsmath,graphicx,multirow}

\newcommand\apj{Astrophys.~J.~}                                          
\newcommand\apjl{Astrophys.~J. Lett.~}                                          
                                         
\newcommand\aap{Astron. Astrophys.~}                                            
                                   
\newcommand\araa{Ann. Rev. Astr. Ap.~}

\newcommand\nat{Nat.~}

\def\plotone#1{\centering \leavevmode
\includegraphics[clip=, width=.85\columnwidth]{#1}}

\newcommand{\cN}[1]{\mathcal{N}}

\def\gsim{\;\rlap{\lower 2.5pt
 \hbox{$\sim$}}\raise 1.5pt\hbox{$>$}\;}
\def\lsim{\;\rlap{\lower 2.5pt
   \hbox{$\sim$}}\raise 1.5pt\hbox{$<$}\;}

\contributor{Submitted to Proceedings
of the National Academy of Sciences of the United States of America}
\url{www.pnas.org/cgi/doi/10.1073/pnas.0709640104}
\copyrightyear{2011}
\issuedate{Issue Date}
\volume{Volume}
\issuenumber{Issue Number}

\begin{document}


\title{Bayesian analysis of the astrobiological implications of
  life's early emergence on Earth}

\author{
David~S.~Spiegel\affil{1}{Institute for Advanced Study, Princeton, NJ
  08540}\affil{2}{Dept. of Astrophysical Sciences, Princeton Univ.,
  Princeton, NJ 08544, USA}\footnotetext{dave@ias.edu},
Edwin~L.~Turner\affil{2}{}\affil{3}{Institute for the Physics and
  Mathematics of the Universe, The Univ. of Tokyo, Kashiwa
  227-8568, Japan}}

\footlineauthor{Spiegel \& Turner}

\contributor{Submitted to Proceedings of the National Academy of
  Sciences of the United States of America}

\maketitle

\begin{article}

\begin{abstract}
Life arose on Earth sometime in the first few hundred million years
after the young planet had cooled to the point that it could support
water-based organisms on its surface.  The early emergence of life on
Earth has been taken as evidence that the probability of abiogenesis
is high, if starting from young-Earth-like conditions.  We revisit
this argument quantitatively in a Bayesian statistical framework.  By
constructing a simple model of the probability of abiogenesis, we
calculate a Bayesian estimate of its posterior probability, given the
data that life emerged fairly early in Earth's history and that,
billions of years later, curious creatures noted this fact and
considered its implications.  We find that, given only this very
limited empirical information, the choice of Bayesian prior for the
abiogenesis probability parameter has a dominant influence on the
computed posterior probability.  Although terrestrial life's early
emergence provides evidence that life might be common in the Universe
if early-Earth-like conditions are, the evidence is inconclusive and
indeed is consistent with an arbitrarily low intrinsic probability of
abiogenesis for plausible uninformative priors.  Finding a single case
of life arising independently of our lineage (on Earth, elsewhere in
the Solar System, or on an extrasolar planet) would provide much
stronger evidence that abiogenesis is not extremely rare in the
Universe.
\end{abstract}


\abbreviations{Gyr, gigayear ($10^9$ years); PDF, probability density
  function; CDF, cumulative distribution function}

\section{Introduction}
\label{sec:intro}
Astrobiology is fundamentally concerned with whether extraterrestrial
life exists and, if so, how abundant it is in the Universe.  The most
direct and promising approach to answering these questions is surely
empirical, the search for life on other bodies in the Solar System
\cite{chyba+hand2005, desmarais+walter1999} and beyond in other
planetary systems \cite{desmarais_et_al2002, seager_et_al2005}.
Nevertheless, a theoretical approach is possible in principle and
could provide a useful complement to the more direct lines of
investigation.

In particular, if we knew the probability per unit time and per unit
volume of abiogenesis in a pre-biotic environment as a function of its
physical and chemical conditions and if we could determine or estimate
the prevalence of such environments in the Universe, we could make a
statistical estimate of the abundance of extraterrestrial life.  This
relatively straightforward approach is, of course, thwarted by our
great ignorance regarding both inputs to the argument at present.

There does, however, appear to be one possible way of finessing our
lack of detailed knowledge concerning both the process of abiogenesis
and the occurrence of suitable pre-biotic environments (whatever they
might be) in the Universe.  Namely, we can try to use our knowledge
that life arose at least once in an environment (whatever it was) on
the early Earth to try to infer something about the probability per
unit time of abiogenesis on an Earth-like planet without the need (or
ability) to say how Earth-like it need be or in what ways.  We will
hereinafter refer to this probability per unit time, which can also be
considered a rate, as $\lambda$ or simply the ``probability of
abiogenesis.''

Any inferences about the probability of life arising (given the
conditions present on the early Earth) must be informed by how long it
took for the first living creatures to evolve.  By definition,
improbable events generally happen infrequently.  It follows that the
duration between events provides a metric (however imperfect) of the
probability or rate of the events.  The time-span between when Earth
achieved pre-biotic conditions suitable for abiogenesis plus generally
habitable climatic conditions \cite{kasting_et_al1993,
  selsis_et_al2007, spiegel_et_al2008} and when life first arose,
therefore, seems to serve as a basis for estimating $\lambda$.
Revisiting and quantifying this analysis is the subject of this paper.

We note several previous quantitative attempts to address this issue
in the literature, of which one \cite{carter1983} found, as we do,
that early abiogenesis is consistent with life being rare, and the
other \cite{lineweaver+davis2002} found that Earth's early abiogenesis
points strongly to life being common on Earth-like planets (we compare
our approach to the problem to that of \cite{lineweaver+davis2002}
below, including our significantly different results).\footnote{There
  are two unpublished works (\cite{brewer2008} and
  \cite{korpela2011}), of which we became aware after submission of
  this paper, that also conclude that early life on Earth does not
  rule out the possibility that abiogenesis is improbable.}
Furthermore, an argument of this general sort has been widely used in
a qualitative and even intuitive way to conclude that $\lambda$ is
unlikely to be extremely small because it would then be surprising for
abiogenesis to have occurred as quickly as it did on Earth
\cite{vanzuilen_et_al2002, westall2005, moorbath2005,
  nutman+friend2006, buick2007, sullivan+baross2007,
  sugitani_et_al2010}.  Indeed, the early emergence of life on Earth
is often taken as significant supporting evidence for ``optimism"
about the existence of extra-terrestrial life ({\it i.e.}, for the
view that it is fairly common) \cite{ward+brownlee2000, darling2001,
  lineweaver+davis2002}.  The major motivation of this paper is to
determine the quantitative validity of this inference.  We emphasize
that our goal is {\it not} to derive an optimum estimate of $\lambda$
based on all of the many lines of available evidence, but simply to
evaluate the implication of life's early emergence on Earth for the
value of $\lambda$.

\section{A Bayesian Formulation of the Calculation}
\label{sec:model}
Bayes's theorem \cite{bayes1763} can be written as
${\rm P}[\mathcal{M} | \mathcal{D}] = \left({\rm P}[\mathcal{D} | \mathcal{M}] \times {\rm P}_{\rm prior}[\mathcal{M}]\right)/{\rm P}[\mathcal{D}]$.
Here, we take $\mathcal{M}$ to be a model and $\mathcal{D}$ to be
data.  In order to use this equation to evaluate the posterior
probability of abiogenesis, we must specify appropriate $\mathcal{M}$
and $\mathcal{D}$.

\subsection{A Poisson or Uniform Rate Model}
\label{ssec:Poisson}
In considering the development of life on a planet, we suggest that a
reasonable, if simplistic, model is that it is a Poisson process
during a period of time from $t_{\rm min}$ until $t_{\rm max}$.
In this model, the conditions on a young planet preclude the
development of life for a time period of $t_{\rm min}$ after its
formation.  Furthermore, if the planet remains lifeless until $t_{\rm
  max}$ has elapsed, it will remain lifeless thereafter as well
because conditions no longer permit life to arise.  For a planet
around a solar-type star, $t_{\rm max}$ is almost certainly
$\lsim$10~Gyr (10 billion years, the main sequence lifetime of the
Sun) and could easily be a substantially shorter period of time if
there is something about the conditions on a young planet that are
necessary for abiogenesis.  Between these limiting times, we posit
that there is a certain probability per unit time ($\lambda$) of life
developing.  For $t_{\rm min} < t < t_{\rm max}$, then, the
probability of life arising $n$ times in time $t$ is
\begin{equation}
{\rm P}[\lambda,n,t] = {\rm P}_{\rm Poisson}[\lambda,n,t] = e^{-\lambda (t-t_{\rm min})} \frac{\{ \lambda (t-t_{\rm min})\}^n}{n!} \, ,
\label{eq:poisson}
\end{equation}
where $t$ is the time since the formation of the planet.  

This formulation could well be questioned on a number of grounds.
Perhaps most fundamentally, it treats abiogenesis as though it were a
single instantaneous event and implicitly assumes that it can occur in
only a single way ({\it i.e.}, by only a single process or mechanism)
and only in one type of physical environment.  It is, of course, far
more plausible that abiogenesis is actually the result of a complex
chain of events that take place over some substantial period of time
and perhaps via different pathways and in different environments.
However, knowledge of the actual origin of life on Earth, to say
nothing of other possible ways in which it might originate, is so
limited that a more complex model is not yet justified.  In essence,
the simple Poisson event model used in this paper attempts to
``integrate out'' all such details and treat abiogenesis as a ``black
box'' process: certain chemical and physical conditions as input
produce a certain probability of life emerging as an output.  Another
issue is that $\lambda$, the probability per unit time, could itself
be a function of time.  In fact, the claim that life could not have
arisen outside the window $(t_{\rm min},t_{\rm max})$ is tantamount to
saying that $\lambda=0$ for $t\le t_{\rm min}$ and for $t\ge t_{\rm
  max}$.  Instead of switching from 0 to a fixed value
instantaneously, $\lambda$ could exhibit a complicated variation with
time.  If so, however, ${\rm P}[\lambda,n,t]$ is not represented by
the Poisson distribution and eq.~(\ref{eq:poisson}) is not valid.
Unless a particular (non top-hat-function) time-variation of $\lambda$
is suggested on theoretical grounds, it seems unwise to add such
unconstrained complexity.

A further criticism is that $\lambda$ could be a function of $n$: it
could be that life arising once (or more) changes the probability per
unit time of life arising again.  Since we are primarily interested in
the probability of life arising {\it at all} -- i.e., the probability
of $n \ne 0$ -- we can define $\lambda$ simply to be the value
appropriate for a prebiotic planet (whatever that value may be) and
remain agnostic as to whether it differs for $n \ge 1$.  Thus, within
the adopted model, the probability of life arising is one minus the
probability of it not arising:
\begin{equation}
{\rm P}_{\rm life} = 1-{\rm P}_{\rm Poisson}[\lambda,0,t] = 1 - e^{-\lambda
  (t-t_{\rm min})} \, .
\label{eq:Plife}
\end{equation}

\subsection{A Minimum Evolutionary Time Constraint}
\label{ssec:evoltime}
Naively, the single datum informing our calculation of the posterior
of $\lambda$ appears to be simply that life arose on Earth at least
once, approximately 3.8 billion years ago (give or take a few hundred
million years).  There is additional significant context for this
datum, however.  Recall that the standard claim is that, since life
arose early on the only habitable planet that we have examined for
inhabitants, the probability of abiogenesis is probably high (in our
language, $\lambda$ is probably large).  This standard argument
neglects a potentially important selection effect, namely: On Earth,
it took nearly 4~Gyr for evolution to lead to organisms capable of
pondering the probability of life elsewhere in the Universe.  If this
is a necessary duration, then it would be impossible for us to find
ourselves on, for example, a ($\sim$4.5-Gyr old) planet on which life
first arose only after the passage of 3.5 billion years
\cite{lineweaver+davis2003}.  On such planets there would not yet have
been enough time for creatures capable of such contemplations to
evolve.  In other words, if evolution {\it requires} 3.5~Gyr for life
to evolve from the simplest forms to intelligent, questioning beings,
then we {\it had} to find ourselves on a planet where life arose
relatively early, regardless of the value of $\lambda$.

In order to introduce this constraint into the calculation we define
$\delta t_{\rm evolve}$ as the minimum amount of time required after
the emergence of life for cosmologically curious creatures to evolve,
$t_{\rm emerge}$ as the age of the Earth from when the earliest extant
evidence of life remains (though life might have actually emerged
earlier), and $t_0$ as the current age of the Earth.  The data, then,
are that life arose on Earth at least once, approximately 3.8 billion
years ago, {\it and} that this emergence was early enough that human
beings had the opportunity subsequently to evolve and to wonder about
their origins and the possibility of life elsewhere in the Universe.
In equation form, $t_{\rm emerge} < t_0 - \delta t_{\rm evolve}$.

\begin{table}[t]
\small
\vspace{0.2in}
\label{ta:models}
{\it Models of $t_0=4.5$~Gyr-Old Planets\\}
\begin{tabular}{l|cccc}
\hline
\hline
              Model   & {\tt Hypothetical}  &  {\tt Conserv.$_1$} &  {\tt Conserv.$_2$} & {\tt Optimistic}    \\
\hline
\rule {-3pt} {10pt}
$t_{\rm min}$        & 0.5                & 0.5                    & 0.5                     & 0.5                \\[0.2cm]
$t_{\rm emerge}$         & 0.51               & 1.3                    & 1.3                     & 0.7                \\[0.2cm]
$t_{\rm max}$           & 10                 & 1.4                    & 10                      & 10                  \\[0.2cm]
$\delta t_{\rm evolve}$  & 1                  & 2                     & 3.1                      & 1                  \\[0.2cm]
$t_{\rm required}$            & 3.5                & 1.4                   & 1.4                      & 3.5                \\[0.2cm]
$\Delta t_1$            & 0.01               & 0.80                  & 0.80                     & 0.20               \\[0.2cm]
$\Delta t_2$            & 3.00               & 0.90                  & 0.90                     & 3.00               \\[0.2cm]
$\mathcal{R}$           & 300                & 1.1                   & 1.1                      & 15                 \\[0.2cm]
\hline
\end{tabular}
\begin{flushleft}
{\small All times are in Gyr.  Two ``Conservative'' ({\tt
    Conserv.}) models are shown, to indicate that $t_{\rm required}$ may be
  limited either by a small value of $t_{\rm max}$ (``{\tt
    Conserv.$_1$}''), or by a large value of $\delta t_{\rm evolve}$
  (``{\tt Conserv.$_2$}'').}
\end{flushleft}
\end{table}

\subsection{The Likelihood Term}
\label{ssec:likelihood}
We now seek to evaluate the ${\rm P}[\mathcal{D}|\mathcal{M}]$ term in
Bayes's theorem.  Let $t_{\rm required} \equiv \min[t_0-\delta
  t_{\rm evolve},t_{\rm max}]$.  Our existence on Earth requires that
life appeared within $t_{\rm required}$.  In other words, $t_{\rm
  required}$ is the maximum age that the Earth could have had at the
origin of life in order for humanity to have a chance of showing up by
the present.  We define $\mathcal{S_E}$ to be the set of all
Earth-like worlds of age approximately $t_0$ in a large, unbiased
volume and $L[t]$ to be the subset of $\mathcal{S_E}$ on which life
has emerged within a time $t$.  $L[t_{\rm required}]$ is the set of
planets on which life emerged early enough that creatures curious
about abiogenesis could have evolved before the present ($t_0$), and,
presuming $t_{\rm emerge} < t_{\rm required}$ (which we know was the
case for Earth), $L[t_{\rm emerge}]$ is the subset of $L[t_{\rm
    required}]$ on which life emerged as quickly as it did on Earth.
Correspondingly, $N_\mathcal{S_E}$, $N_{t_{\rm r}}$, and $N_{t_{\rm
    e}}$ are the respective numbers of planets in sets
$\mathcal{S_E}$, $L[t_{\rm required}]$, and $L[t_{\rm emerge}]$.  The
fractions $\varphi_{t_{\rm r}} \equiv N_{t_{\rm r}}/N_\mathcal{S_E}$
and $\varphi_{t_{\rm e}} \equiv N_{t_{\rm e}}/N_\mathcal{S_E}$ are,
respectively, the fraction of Earth-like planets on which life arose
within $t_{\rm required}$ and the fraction on which life emerged
within $t_{\rm emerge}$.  The ratio $r \equiv \varphi_{t_{\rm e}} /
\varphi_{t_{\rm r}} = N_{t_{\rm e}} / N_{t_{\rm r}}$ is the fraction
of $L_{t_{\rm r}}$ on which life arose as soon as it did on Earth.
Given that we {\it had to} find ourselves on such a planet in the set
$L_{t_{\rm r}}$ in order to write and read about this topic, the ratio
$r$ characterizes the probability of the data given the model if the
probability of intelligent observers arising is independent of the
time of abiogenesis (so long as abiogenesis occurs before $t_{\rm
  required}$).  (This last assumption might seem strange or
unwarranted, but the effect of relaxing this assumption is to make it
more likely that we would find ourselves on a planet with early
abiogenesis and therefore to reduce our limited ability to infer
anything about $\lambda$ from our observations.) Since
$\varphi_{t_{\rm e}} = 1 - {\rm P}_{\rm Poisson}[\lambda,0,t_{\rm
    emerge}]$ and $\varphi_{t_{\rm r}} = 1 - {\rm P}_{\rm
  Poisson}[\lambda,0,t_{\rm required}]$, we may write that
\begin{equation}
{\rm P}[\mathcal{D}|\mathcal{M}] = \frac{1 - \exp[-\lambda (t_{\rm emerge}-t_{\rm min})]}{1 - \exp[-\lambda (t_{\rm required}-t_{\rm min})]}
\label{eq:Pdata}
\end{equation}
if $t_{\rm min} < t_{\rm emerge} < t_{\rm required}$ (and ${\rm
  P}[\mathcal{D}|\mathcal{M}]=0$ otherwise).  This is called the
``likelihood function,'' and represents the probability of the
observation(s), given a particular model.\footnote{An alternative way
  to derive equation~(\ref{eq:Pdata}) is to let $E$ = ``abiogenesis
  occurred between $t_{\rm min}$ and $t_{\rm emerge}$'' and $R$ =
  ``abiogenesis occurred between $t_{\rm min}$ and $t_{\rm
    required}$.''  We then have, from the rules of conditional probability,
  ${\rm P}[E|R,\mathcal{M}] = {\rm P}[E,R|\mathcal{M}]/{\rm
    P}[R|\mathcal{M}]$.  Since $E$ entails $R$, the numerator on the
  right-hand side is simply equal to ${\rm P}[E|\mathcal{M}]$, which
  means that the previous equation reduces to
  equation~(\ref{eq:Pdata}).} It is via this function that the data
``condition'' our prior beliefs about $\lambda$ in standard Bayesian
terminology.

\subsection{Limiting Behavior of the Likelihood}
\label{ssec:limiting}
It is instructive to consider the behavior of
equation~(\ref{eq:Pdata}) in some interesting limits.  For $\lambda
(t_{\rm required} - t_{\rm min}) \ll 1$, the numerator and
denominator of equation~(\ref{eq:Pdata}) each go approximately as the
argument of the exponential function; therefore, in this limit, the
likelihood function is approximately constant:
\begin{equation}
{\rm P}[\mathcal{D}|\mathcal{M}] \approx \frac{t_{\rm emerge} - t_{\rm min}}{t_{\rm required} - t_{\rm min}} \, .
\label{eq:smalllim}
\end{equation}
This result is intuitively easy to understand as follows: If $\lambda$
is sufficiently small, it is overwhelmingly likely that abiogenesis
occurred only once in the history of the Earth, and by the assumptions
of our model, the one event is equally likely to occur at any time
during the interval between $t_{\rm min}$ and $t_{\rm required}$.  The
chance that this will occur by $t_{\rm emerge}$ is then just the
fraction of that total interval that has passed by $t_{\rm emerge}$ --
the result given in equation~(\ref{eq:smalllim}).

In the other limit, when $\lambda (t_{\rm emerge} - t_{\rm
  min}) \gg 1$, the numerator and denominator of
equation~(\ref{eq:Pdata}) are both approximately 1.  In this case, the
likelihood function is also approximately constant (and equal to
unity).  This result is even more intuitively obvious since a very
large value of $\lambda$ implies that abiogenesis events occur at a
high rate (given suitable conditions) and are thus likely to have
occurred very early in the interval between $t_{\rm min}$ and
$t_{\rm required}$.

These two limiting cases, then, already reveal a key conclusion of our
analysis: the posterior distribution of $\lambda$ for both very large
and very small values will have the shape of the prior, just scaled by
different constants.  Only when $\lambda$ is neither very large nor
very small -- or, more precisely, when $\lambda (t_{\rm emerge} -
t_{\rm min}) \approx 1$ -- do the data and the prior both inform the
posterior probability at a roughly equal level.

\subsection{The Bayes Factor}
\label{ssec:bfactor}
In this context, note that the probability in
equation~(\ref{eq:Pdata}) depends crucially on two time differences,
$\Delta t_1 \equiv t_{\rm emerge} - t_{\rm min}$ and $\Delta
t_2 \equiv t_{\rm required} - t_{\rm min}$, and that the ratio of
the likelihood function at large $\lambda$ to its value at small
$\lambda$ goes roughly as
\begin{equation}
\mathcal{R} \equiv \frac{{\rm P}[\rm data|large~\lambda]}{{\rm P}[\rm data|small~\lambda]} \approx \frac{\Delta t_2}{\Delta t_1} \, .
\label{eq:ratio}
\end{equation}
$\mathcal{R}$ is called the Bayes factor or Bayes ratio and is
sometimes employed for model selection purposes.  In one conventional
interpretation \cite{jeffreys1961}, $\mathcal{R} \le 10$ implies no
strong reason {\it in the data alone} to prefer the model in the
numerator over the one in the denominator.  For the problem at hand,
this means that the datum does not justify preference for a large
value of $\lambda$ over an arbitrarily small one unless
equation~(\ref{eq:ratio}) gives a result larger than roughly ten.

Since the likelihood function contains all of the information in the
data and since the Bayes factor has the limiting behavior given in
equation~\ref{eq:ratio}, our analysis in principle need not consider
priors.  If a small value of $\lambda$ is to be decisively ruled out
by the data, the value of $\mathcal{R}$ must be much larger than
unity.  It is not for plausible choices of the parameters (see
Table~1), and thus arbitrarily small values of $\lambda$ can only be
excluded by some adopted prior on its values.  Still, for illustrative
purposes, we now proceed to demonstrate the influence of various
possible $\lambda$ priors on the $\lambda$ posterior.

\begin{figure}[t]
\begin{center}
\includegraphics[width=8.cm,angle=0,clip=true]{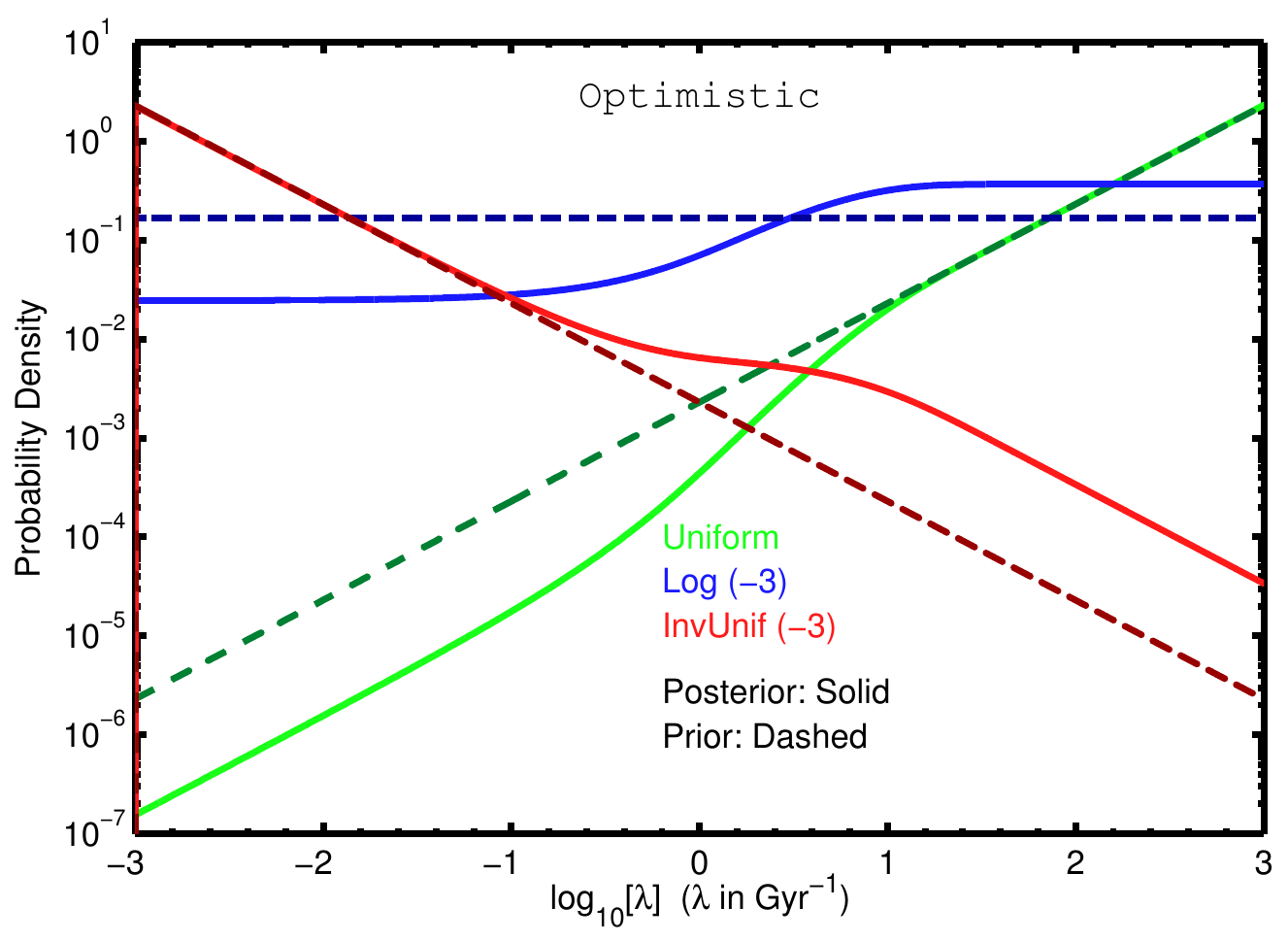}\\
\includegraphics[width=8.cm,angle=0,clip=true]{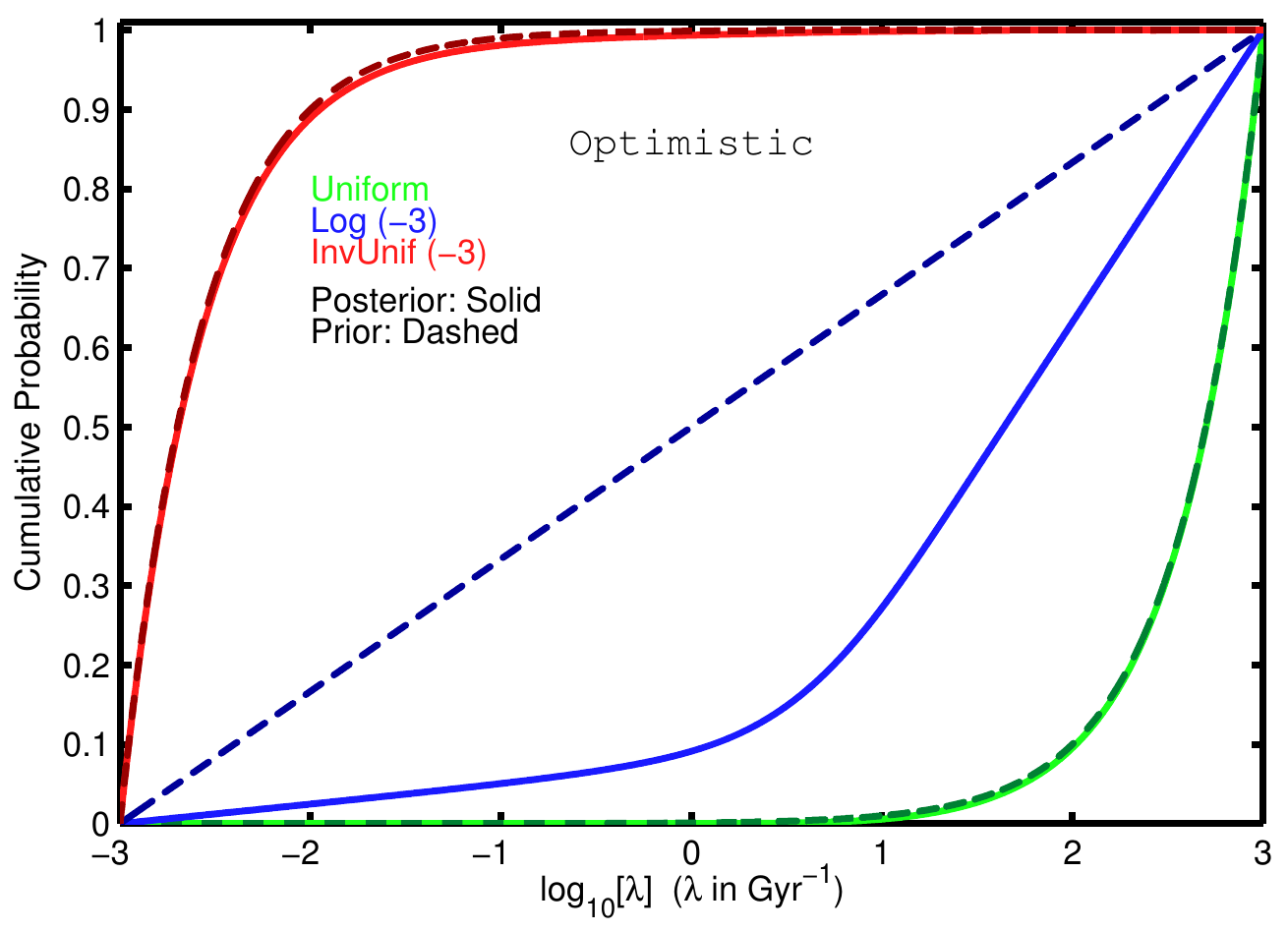}
\end{center}
\caption{\small PDF and CDF of $\lambda$ for uniform, logarithmic, and
  inverse-uniform priors, for model {\tt Optimistic}, with
  $\lambda_{\rm min} = 10^{-3} \rm~Gyr^{-1}$ and $\lambda_{\rm max} =
  10^3 \rm~Gyr^{-1}$.  {\it Top:} The dashed and solid curves
  represent, respectively, the prior and posterior probability
  distribution functions (PDFs) of $\lambda$ under three different
  assumptions about the nature of the prior.  The green curves are for
  a prior that is uniform on the range $0 {\rm~Gyr^{-1}} \le \lambda
  \le \lambda_{\rm max}$ (``Uniform''); the blue are for a prior that
  is uniform in the log of $\lambda$ on the range $-3 \le \log \lambda
  \le 3$ (``Log (-3)''); and the red are for a prior that is uniform
  in $\lambda^{-1}$ on the interval $10^{-3} {\rm~Gyr} \le
  \lambda^{-1} \le 10^3 {\rm~Gyr}$ (``InvUnif (-3)'').  {\it Bottom:}
  The curves represent the cumulative distribution functions (CDFs) of
  $\lambda$.  The ordinate on each curve represents the integrated
  probability from 0 to the abscissa (color and line-style schemes are
  the same as in the top panel).  For a uniform prior, the posterior
  CDF traces the prior almost exactly.  In this case, the posterior
  judgment that $\lambda$ is probably large simply reflects the prior
  judgment of the distribution of $\lambda$.  For the prior that is
  uniform in $\lambda^{-1}$ (InvUnif), the posterior judgment is quite
  opposite -- namely, that $\lambda$ is probably quite small -- but
  this judgment is also foretold by the prior, which is traced nearly
  exactly by the posterior.  For the logarithmic prior, the datum
  (that life on Earth arose within a certain time window) does
  influence the posterior assessment of $\lambda$, shifting it in the
  direction of making greater values of $\lambda$ more probable.
  Nevertheless, the posterior probability is $\sim$12\% that $\lambda
  < 1 \rm~Gyr^{-1}$.  Lower $\lambda_{\rm min}$ and/or lower
  $\lambda_{\rm max}$ would further increase the posterior probability
  of very low $\lambda$, for any of the priors.}
\label{fig:pdfcdf_just3}
\end{figure}

\subsection{The Prior Term}
\label{ssec:prior}
To compute the desired posterior probability, what remains to be
specified is ${\rm P}_{\rm prior}[\mathcal{M}]$, the prior joint
probability density function (PDF) of $\lambda$, $t_{\rm min}$,
$t_{\rm max}$, and $\delta t_{\rm evolve}$.
One approach to choosing appropriate priors for $t_{\rm min}$, $t_{\rm
  max}$, and $\delta t_{\rm evolve}$, would be to try to distill
geophysical and paleobiological evidence along with theories for the
evolution of intelligence and the origin of life into quantitative
distribution functions that accurately represent prior information and
beliefs about these parameters.  Then, in order to ultimately
calculate a posterior distribution of $\lambda$, one would marginalize
over these ``nuisance parameters.''  However, since our goal is to
evaluate the influence of life's early emergence on our posterior
judgment of $\lambda$ (and not of the other parameters), we instead
adopt a different approach.  Rather than calculating a posterior over
this 4-dimensional parameter space, we investigate the way these three
time parameters affect our inferences regarding $\lambda$ by simply
taking their priors to be delta functions at several theoretically
interesting values: a purely hypothetical situation in which life
arose extremely quickly, a most conservative situation, and an in
between case that is also optimistic but for which there does exist
some evidence (see Table~1).

For the values in Table~1, the likelihood ratio $\mathcal{R}$ varies
from $\sim$1.1 to 300, with the parameters of the ``optimistic'' model
giving a borderline significance value of $\mathcal{R} = 15$.  Thus,
only the hypothetical case gives a decisive preference for large
$\lambda$ by the Bayes factor metric, and we emphasize that there is
no direct evidence that abiogenesis on Earth occurred that early, only
10 million years after conditions first permitted
it!\footnote{\cite{lazcano+miller1994} advances this claim based on
  theoretical arguments that are critically reevaluated in
  \cite{orgel1998}}

We also lack a first-principles theory or other solid prior
information for $\lambda$.  We therefore take three different
functional forms for the prior -- uniform in $\lambda$, uniform in
$\lambda^{-1}$ (equivalent to saying that the mean time until life
appears is uniformly distributed), and uniform in $\log_{10}\lambda$.
For the uniform in $\lambda$ prior, we take our prior confidence in
$\lambda$ to be uniformly distributed on the interval 0 to
$\lambda_{\rm max} = 1000$ $\rm~Gyr^{-1}$ (and to be 0 otherwise).
For the uniform in $\lambda^{-1}$ and the uniform in
$\log_{10}[\lambda]$ priors, we take the prior density functions for
$\lambda^{-1}$ and $\log_{10}[\lambda]$, respectively, to be uniform
on $\lambda_{\rm min} \le \lambda \le \lambda_{\rm max}$ (and 0
otherwise).  For illustrative purposes, we take three values of
$\lambda_{\rm min}$: $10^{-22} \rm~Gyr^{-1}$, $10^{-11} \rm~Gyr^{-1}$,
and $10^{-3} \rm~Gyr^{-1}$, corresponding roughly to life occuring
once in the observable Universe, once in our galaxy, and once per 200
stars (assuming one Earth-like planet per star).

In standard Bayesian terminology, both the uniform in $\lambda$ and
the uniform in $\lambda^{-1}$ priors are said to be highly
``informative.''  This means that they strongly favor large and small,
respectively, values of $\lambda$ in advance, {\it i.e., on some basis
  other than the empirical evidence represented by the likelihood
  term}.  For example, the uniform in $\lambda$ prior asserts that we
know on some other basis (other than the early emergence of life on
Earth) that it is a hundred times less likely that $\lambda$ is less
than $10^{-3} \rm~Gyr^{-1}$ than that it is less than $0.1
\rm~Gyr^{-1}$.  The uniform in $\lambda^{-1}$ prior has the equivalent
sort of preference for small $\lambda$ values.  By contrast, the
logarithmic prior is relatively ``uninformative'' in standard Bayesian
terminology and is equivalent to asserting that we have no prior
information that informs us of even the order-of-magnitude of
$\lambda$.

In our opinion, the logarithmic prior is the most appropriate one
given our current lack of knowledge of the process(es) of abiogenesis,
as it represents scale-invariant ignorance of the value of $\lambda$.
It is, nevertheless, instructive to carry all three priors through the
calculation of the posterior distribution of $\lambda$, because they
vividly illuminate the extent to which the result depends on the data
vs the assumed prior.

\subsection{Comparison with Previous Analysis}
\label{ssec:Lineweaver}
Using a binomial probability analysis, Lineweaver \& Davis
\cite{lineweaver+davis2002} attempted to quantify $q$, the probability
that life would arise within the first billion years on an Earth-like
planet.  Although the binomial distribution typically applies to
discrete situations (in contrast to the continuous passage of time,
during which life might arise), there is a simple correspondence
between their analysis and the Poisson model described above.  The
probability that life would arise at least once within a billion years
(what \cite{lineweaver+davis2002} call $q$) is a simple transformation
of $\lambda$, obtained from equation~(\ref{eq:Plife}), with $\Delta
t_1 = 1$~Gyr:
\begin{equation}
q = 1 - e^{-(\lambda) (1 \rm~Gyr)} \qquad \mbox{or} \qquad \lambda = \ln[1 - q]/(1 \rm~Gyr) \, .
\label{eq:qlam}
\end{equation}

In the limit of $\lambda (1 {\rm~Gyr}) \ll 1$,
equation~(\ref{eq:qlam}) implies that $q$ is equal to $\lambda (1
{\rm~Gyr})$.  Though not cast in Bayesian terms, the analysis in
\cite{lineweaver+davis2002} draws a Bayesian conclusion and therefore
is based on an implicit prior that is uniform in $q$.  As a result, it
is equivalent to our uniform-$\lambda$ prior for small values of
$\lambda$ (or $q$), and it is this implicit prior, not the early
emergence of life on Earth, that dominates their conclusions.

\section{The Posterior Probability of Abiogenesis}
\label{sec:results}
We compute the normalized product of the probability of the data given
$\lambda$ (equation~\ref{eq:Pdata}) with each of the three priors
(uniform, logarithmic, and inverse uniform).  This gives us the
Bayesian posterior PDF of $\lambda$, which we also derive for each
model in Table~1.  Then, by integrating each PDF from $-\infty$ to
$\lambda$, we obtain the corresponding cumulative distribution
function (CDF).

Figure~1 displays the results by plotting the prior and posterior
probability of $\lambda$.  The top panel presents the PDF, and the
bottom panel the CDF, for uniform, logarithmic, and inverse-uniform
priors, for model {\tt Optimistic}, which sets $\Delta t_1$ (the
maximum time it might have taken life to emerge once Earth became
habitable) to 0.2~Gyr, and $\Delta t_2$ (the time life had available
to emerge in order that intelligent creatures would have a chance to
evolve) to 3.0~Gyr.  The dashed and solid curves represent,
respectively, prior and posterior probability functions.  In this
figure, the priors on $\lambda$ have $\lambda_{\rm min} = 10^{-3}
\rm~Gyr^{-1}$ and $\lambda_{\rm max} = 10^3 \rm~Gyr^{-1}$.  The green,
blue, and red curves are calculated for uniform, logarithmic, and
inverse-uniform priors, respectively.  The results of the
corresponding calculations for the other models and bounds on the
assumed priors are presented in the Supporting Information, but the
cases shown in Fig.~1 suffice to demonstrate all of the important
qualitative behaviors of the posterior.

In the plot of differential probability (PDF; top panel), it appears
that the inferred posterior probabilities of different values of
$\lambda$ are conditioned similarly by the data (leading to a jump in
the posterior PDF of roughly an order of magnitude in the vicinity of
$\lambda \sim 0.5$~Gyr$^{-1}$).  The plot of cumulative probability,
however, immediately shows that the uniform and the inverse priors
produce posterior CDFs that are completely insensitive to the data.
Namely, small values of $\lambda$ are strongly excluded in the uniform
in $\lambda$ prior case and large values are equally strongly excluded
by the uniform in $\lambda^{-1}$ prior, but these strong conclusions
are not a consequence of the data, only of the assumed prior.  This
point is particularly salient, given that a Bayesian interpretation of
\cite{lineweaver+davis2002} indicates an implicit uniform prior.  In
other words, their conclusion that $q$ cannot be too small and thus
that life should not be too rare in the Universe is {\it not} a
consequence of the evidence of the early emergence of life on the
Earth but almost only of their particular parameterization of the
problem.

\begin{figure}[t]
\begin{center}
\includegraphics[width=8.cm,angle=0,clip=true]{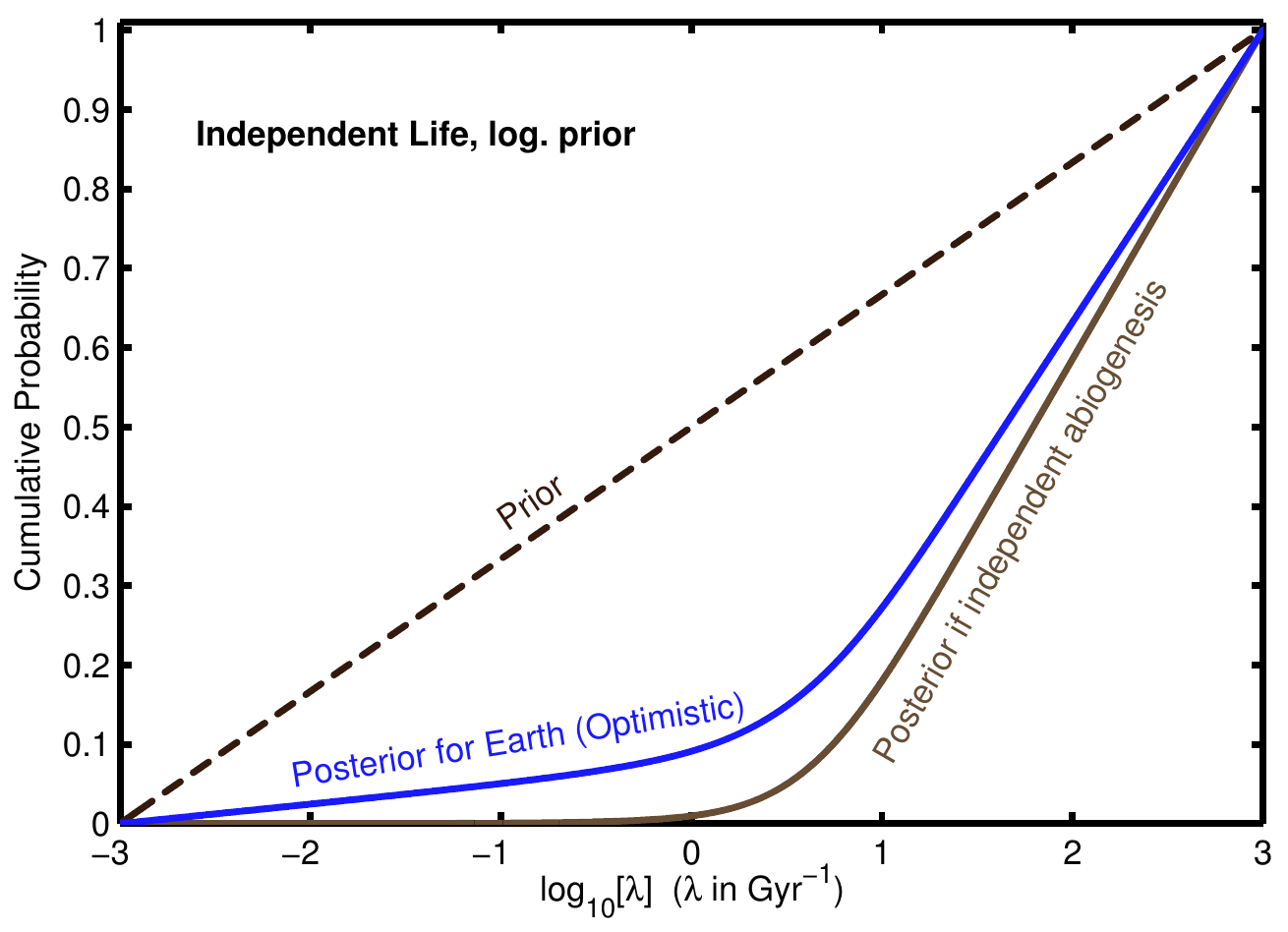}
\end{center}
\caption{CDF of $\lambda$ for abiogenesis with independent lineage,
  for logarithmic prior.  $\lambda_{\rm min} = 10^{-3} \rm~Gyr^{-1}$,
  $\lambda_{\rm max} = 10^3 \rm~Gyr^{-1}$.  A discovery that life
  arose independently on Mars and Earth or on an exoplanet and Earth
  -- or that it arose a second, independent, time on Earth -- would
  significantly reduce the posterior probability of low $\lambda$.}
\label{fig:pdfcdf_mars}
\end{figure}

For the {\tt Optimistic} parameters, the posterior CDF computed with
the uninformative logarithmic prior does reflect the influence of the
data, making greater values of $\lambda$ more probable in accordance
with one's intuitive expectations.  However, with this relatively
uninformative prior, there is a significant probability that $\lambda$
is very small (12\% chance that $\lambda<1\rm~Gyr^{-1}$).  Moreover,
if we adopted smaller $\lambda_{\rm min}$, smaller $\lambda_{\rm
  max}$, and/or a larger $\Delta t_1/\Delta t_2$ ratio, the posterior
probability of an arbitrarily low $\lambda$ value can be made
acceptably high (see Fig.~3 and the Supporting Information).

\subsection{Independent Abiogenesis}
\label{ssec:mars}
We have no strong evidence that life ever arose on Mars (although no
strong evidence to the contrary either).  Recent observations have
tenatively suggested the presence of methane at the level of $\sim$20
parts per billion (ppb) \cite{mumma_et_al2009}, which could
potentially be indicative of biological activity.  The case is not
entirely clear, however, as alternative analysis of the same data
suggests that an upper limit to the methane abundance is in the
vicinity of $\sim$3~ppb \cite{zahnle_et_al2011}.  If, in the future,
researchers find compelling evidence that Mars or an exoplanet hosts
life that arose independently of life on Earth (or that life arose on
Earth a second, independent time \cite{davies+lineweaver2005,
  davies_et_al2009}), how would this affect the posterior probability
density of $\lambda$ (assuming that the same $\lambda$ holds for both
instances of abiogenesis)?

If Mars, for instance, and Earth share a single $\lambda$ and life
arose arise on Mars, then the likelihood of Mars' $\lambda$ is the
joint probability of our data on Earth and of life arising on Mars.
Assuming no panspermia in either direction, these events are
independent:
\begin{eqnarray}
\nonumber {\rm P}[\mathcal{D}|\mathcal{M}] & = & \left(1 - \exp[-\lambda (t_{\rm emerge}^{\rm Mars}-t_{\rm min}^{\rm Mars})]\right) \\
 & & \times \frac{1 - \exp[-\lambda (t_{\rm emerge}^{\rm Earth}-t_{\rm min}^{\rm Earth})]}{1 - \exp[-\lambda (t_{\rm required}^{\rm Earth}-t_{\rm min}^{\rm Earth})]} \, .
\label{eq:PdataMars_life}
\end{eqnarray}

For Mars, we take $t_{\rm max}^{\rm Mars} = t_{\rm emerge}^{\rm Mars}
= 1$~Gyr and $t_{\rm min}^{\rm Mars} = 0.5$~Gyr.  The posterior
cumulative probability distribution of $\lambda$, given a logarithmic
prior between 0.001~Gyr$^{-1}$ and 1000~Gyr$^{-1}$, is as represented
in Fig.~2 for the case of finding a second, independent sample of life
and, for comparison, the {\tt Optimistic} case for Earth.  Should
future researchers find that life arose independently on Mars (or
elsewhere), this would dramatically reduce the posterior probability
of very low $\lambda$ relative to our current inferences.

\subsection{Arbitrarily Low Posterior Probability of $\lambda$}
\label{ssec:lowpost}
We do not actually know what the appropriate lower (or upper) bounds
on $\lambda$ are.  Figure~3 portrays the influence of changing
$\lambda_{\rm min}$ on the median posterior estimate of $\lambda$, and
on 1-$\sigma$ and 2-$\sigma$ confidence lower bounds on posterior
estimates of $\lambda$.  Although the median estimate is relatvely
insensitive to $\lambda_{\rm min}$, a 2-$\sigma$ lower bound on
$\lambda$ becomes arbitrarily low as $\lambda_{\rm min}$ decreases.
\begin{figure}[t]
\includegraphics[width=8.cm,angle=0,clip=true]{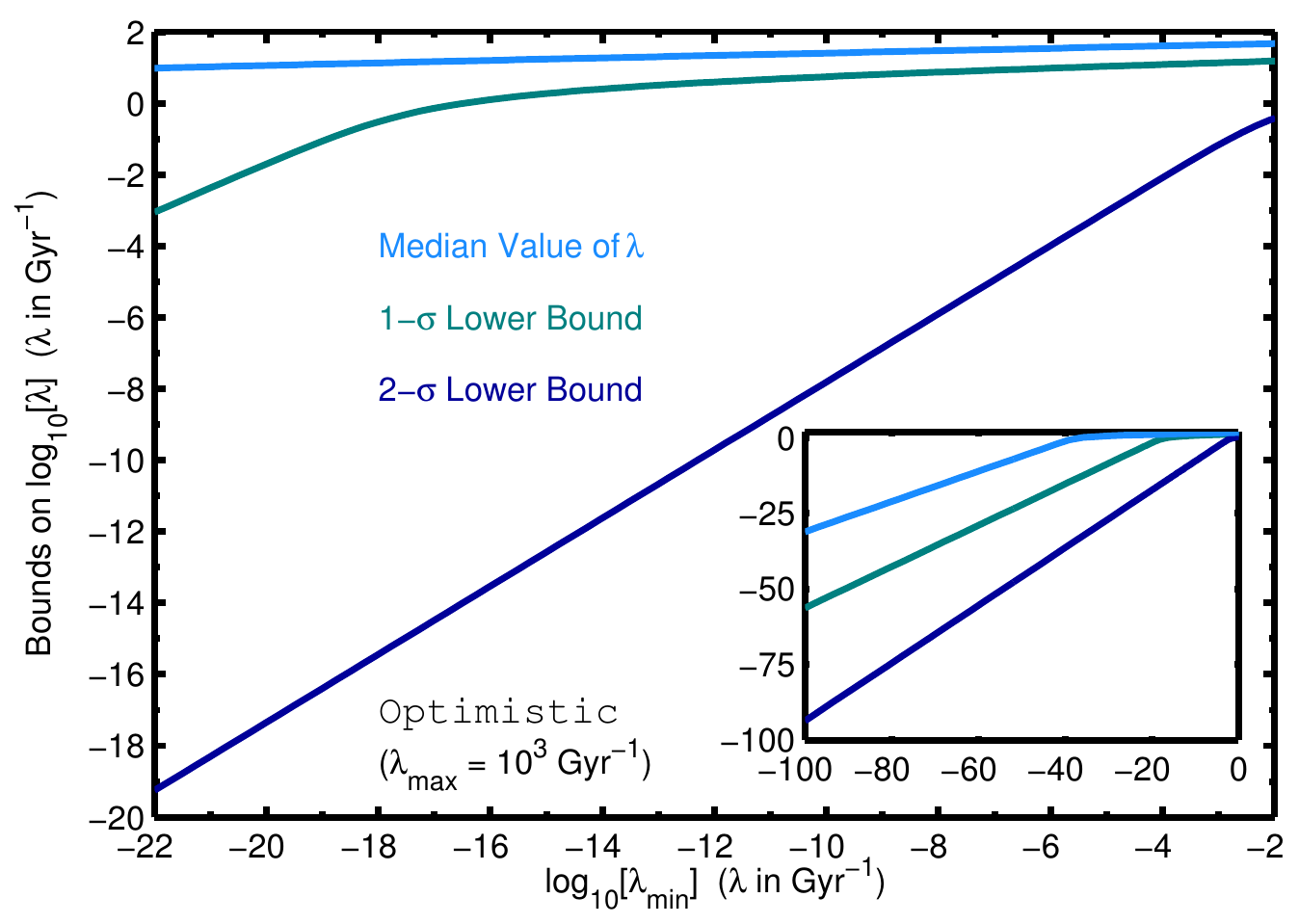}
\caption{Lower bound on $\lambda$ for logarithmic prior, {\tt
    Hypothetical} model.  The three curves depict median (50\%),
  1-$\sigma$ (68.3\%), and 2-$\sigma$ (95.4\%) lower bounds on
  $\lambda$, as a function of $\lambda_{\rm min}$.}
\label{fig:lower_bound}
\end{figure}

\section{Conclusions}
\label{sec:conc}
Within a few hundred million years, and perhaps far more quickly, of
the time that Earth became a hospitable location for life, it
transitioned from being merely habitable to being inhabited.  Recent
rapid progress in exoplanet science suggests that habitable worlds
might be extremely common in our galaxy \cite{vogt_et_al2010,
  borucki_et_al2011, howard_et_al2011, wordsworth_et_al2011}, which
invites the question of how often life arises, given habitable
conditions.  Although this question ultimately must be answered
empirically, via searches for biomarkers \cite{kaltenegger+selsis2007}
or for signs of extraterrestrial technology \cite{tarter2001}, the
early emergence of life on Earth gives us some information about the
probability that abiogenesis will result from early-Earth-like
conditions.\footnote{We note that the comparatively very {\it late}
  emergence of radio technology on Earth could, analogously, be taken
  as an indication (albeit a weak one because of our single datum)
  that radio technology might be rare in our galaxy.}

A Bayesian approach to estimating the probability of abiogenesis
clarifies the relative influence of data and of our prior beliefs.
Although a ``best guess'' of the probability of abiogenesis suggests
that life should be common in the Galaxy if early-Earth-like
conditions are, still, the data are consistent (under plausible
priors) with life being extremely rare, as shown in Figure~3.  Thus, a
Bayesian enthusiast of extraterrestrial life should be significantly
encouraged by the rapid appearance of life on the early Earth but
cannot be highly confident on that basis.

Our conclusion that the early emergence of life on Earth is consistent
with life being very rare in the Universe for plausible priors is
robust against two of the more fundamental simplifications in our
formal analysis.  First, we have assumed that there is a single value
of $\lambda$ that applies to all Earth-like planets (without
specifying exactly what we mean by ``Earth-like'').  If $\lambda$
actually varies from planet to planet, as seems far more plausible,
anthropic-like considerations imply planets with particularly large
$\lambda$ values will have a greater chance of producing (intelligent)
life and of life appearing relatively rapidly, {\it i.e.}, of the
circumstances in which we find ourselves.  Thus, the information we
derive about $\lambda$ from the existence and early appearance of life
on Earth will tend to be biased towards large values and may not be
representative of the value of $\lambda$ for, say, an ``average''
terrestrial planet orbiting within the habitable zone of a main
sequence star.  Second, our formulation of the problem analyzed in
this paper implicitly assumes that there is no increase in the
probability of intelligent life appearing once $\delta t_{\rm evolve}$
has elapsed following the abiogenesis event on a planet.  A more
reasonable model in which this probability continues to increase as
additional time passes would have the same qualitative effect on the
calculation as increasing $\delta t_{\rm evolve}$.  In other words, it
would make the resulting posterior distribution of $\lambda$ even less
sensitive to the data and more highly dependent on the prior because
it would make our presence on Earth a selection bias favoring planets
on which abiogenesis occurred quickly.

We had to find ourselves on a planet that has life on it, but we did
not have to find ourselves (\emph{i}) in a galaxy that has life on a
planet besides Earth nor (\emph{ii}) on a planet on which life arose
multiple, independent times.  Learning that either (\emph{i}) or
(\emph{ii}) describes our world would constitute data that are not
subject to the selection effect described above.  In short, if we
should find evidence of life that arose wholly idependently of us --
either via astronomical searches that reveal life on another planet or
via geological and biological studies that find evidence of life on
Earth with a different origin from us -- we would have considerably
stronger grounds to conclude that life is probably common in our
galaxy.  With this in mind, research in the fields of astrobiology and
origin of life studies might, in the near future, help us to
significantly refine our estimate of the probability (per unit time,
per Earth-like planet) of abiogenesis.

\begin{acknowledgments}
We thank Adam Brown, Adam Burrows, Chris Chyba, Scott Gaudi, Aaron
Goldman, Alan Guth, Larry Guth, Laura Landweber, Tullis Onstott, Caleb
Scharf, Stanley Spiegel, Josh Winn, and Neil Zimmerman for thoughtful
discussions.  ELT is grateful to Carl Boettiger for calling this
problem to his attention some years ago.  DSS acknowledges support
from NASA grant NNX07AG80G, from JPL/Spitzer Agreements 1328092,
1348668, and 1312647, and gratefully acknowledges support from NSF
grant AST-0807444 and the Keck Fellowship.  ELT gratefully
acknowledges support from the World Premier International Research
Center Initiative (Kavli-IPMU), MEXT, Japan and the Research Center for the
Early Universe at the University of Tokyo as well as the hospitality
of its Department of Physics.  Finally, we thank two anonymous
referees for comments that materially improved this manuscript.
\end{acknowledgments}


\begin{onecolumn}

\section*{Supplementary Material}
\label{sec:supp}

\subsection*{Formal Derivation of the Posterior Probability of Abiogenesis}
\label{ssec:deriv}

Let $t_a$ be the time of abiogenesis and $t_i$ be the time of the
emergence of intelligence ($t_0$, $t_{\rm emerge}$, and $t_{\rm required}$
are as defined in the text: $t_0$ is the current age of the Earth;
$t_{\rm emerge}$ is the upper limit on the age of the Earth when life
first arose; and $t_{\rm required}$ is the maximum age the Earth could have
had when life arose in order for it to be possible for sentient beings
to later arise by $t_0$).  By ``intelligence'', we mean organisms that
think about abiogenesis.  Furthermore, let\\
$E = t_{\rm min} < t_a < t_{\rm emerge}$\\
$R = t_{\rm min} < t_a < t_{\rm required}$\\
$I = t_{\rm min} < t_i \le t_0$\\
$\mathcal{M} = ``$The Poisson rate parameter has value $\lambda''$

We assert (perhaps somewhat unreasonably) that the probability of
intelligence arising ($I$) is independent of the actual time of
abiogeneis ($t_a$), so long as life shows up within $t_{\rm required}$
($R$).
\begin{equation}
{\rm P}[I|R,\mathcal{M},t_a] = {\rm P}[I|R,\mathcal{M}] \, .
\label{eq:assert}
\end{equation}
Although the probability of intelligence arising could very well be
greater if abiogenesis occurs earlier on a world, the consequence of
relaxing this assertion (discussed in the Conclusion and elsewhere in
the text) is to increase the posterior probability of arbitrarily low
$\lambda$.

Using the conditional version of Bayes's theorem,
\begin{equation}
{\rm P}[t_a|R,\mathcal{M},I] = \frac{{\rm P}[I|R,\mathcal{M},t_a] \times {\rm P}[t_a|R,\mathcal{M}]}{{\rm P}[I|R,\mathcal{M}]} \, ,
\end{equation}
and Eq.~(\ref{eq:assert}) then implies that ${\rm
  P}[t_a|R,\mathcal{M},I] = {\rm P}[t_a|R,\mathcal{M}]$.  An immediate
result of this is that
\begin{equation}
{\rm P}[E | R, \mathcal{M}, I] = {\rm P}[E | R, \mathcal{M}] \, .
\label{eq:assert2}
\end{equation}

We now apply Bayes's theorem again to get the posterior probability of
$\lambda$, given our circumstances and our observations:
\begin{equation}
{\rm P}[\mathcal{M}|E,R,I] = \frac{{\rm P}[E|R,\mathcal{M},I] \times {\rm P}[\mathcal{M}|R,I]}{{\rm P}[E,R,I]}
\label{eq:post}
\end{equation}
Note that, since $t_e<t_r$, $E \Rightarrow R$.  And, as discussed in
the main text, ${\rm P}[E|R,\mathcal{M}] = {\rm P}[E|\mathcal{M}]/{\rm
  P}[R|\mathcal{M}]$.  Finally, since we had to find ourselves on a
planet on which $R$ and $I$ hold, these conditions tell us nothing
about the value of $\lambda$.  In other words, ${\rm
  P}[\mathcal{M}|R,I] = {\rm P}[\mathcal{M}]$.  We therefore use
Eq.~(\ref{eq:assert2}) to rewrite Eq.~(\ref{eq:post}) as the posterior
probability implied in the text:
\begin{equation}
{\rm P}[\mathcal{M}|E,I] = \frac{\frac{{\rm P}[E|\mathcal{M}]}{{\rm P}[R|\mathcal{M}]} \times {\rm P}[\mathcal{M}]}{{\rm P}[E,I]} \, .
\end{equation}

\subsection*{Model-Dependence of Posterior Probability of Abiogenesis}
\label{ssec:dependence}

In the main text, we demonstrated the strong dependence of the
posterior probability of life on the form of the prior for $\lambda$.
Here, we present a suite of additional calculations, for different
bounds to $\lambda$ and for different values of $\Delta t_1$ and
$\Delta t_2$.

Figure~4 displays the results of analogous calculations to those of
Fig.~1, for three sets model of parameters ({\tt Hypothetical}, {\tt
  Optimistic}, {\tt Conservative}) and for three values of
$\lambda_{\rm min}$ ($10^{-22}\rm~Gyr^{-1}$, $10^{-11}\rm~Gyr^{-1}$,
$10^{-3}\rm~Gyr^{-1}$).  For all three models, the posterior CDFs for
the uniform and the inverse-uniform priors almost exactly match the
prior CDFs, and, hence, are almost completely insensitive to the data.
For the {\tt Conservative} model (in which $\Delta t_1 = 0.8$~Gyr and
$\Delta t_2 = 0.9$~Gyr -- certainly not ruled out by available data),
even the logarithmic prior's CDF is barely sensitive to the
observation that there is life on Earth.

Finally, the effect of $\delta t_{\rm evolve}$ -- the minimum
timescale required for sentience to evolve -- is to impose a selection
effect that becomes progressively more severe as $\delta t_{\rm
  evolve}$ approaches $t_0 - t_{\rm emerge}$.  Figure~5 makes this
point vividly.  For the {\tt Optimistic} model, posterior
probabilities are shown as color maps as functions of $\lambda$
(abscissa) and $\delta t_{\rm evolve}$ (ordinate).  At each horizontal
cut across the PDF plots (left column), the values integrate to unity,
as expected for a proper probability density function.  For short
values of $\delta t_{\rm evolve}$, the selection effect (that
intelligent creatures take some time to evolve) is unimportant, and
the data might be somewhat informative about the true distribution of
$\lambda$.  For larger values of $\delta t_{\rm evolve}$, the
selection effect becomes more important, to the point that the
probability of the data given $\lambda$ approaches 1, and the
posterior probability approaches the prior.

\newpage

\begin{figure}
\plotone
{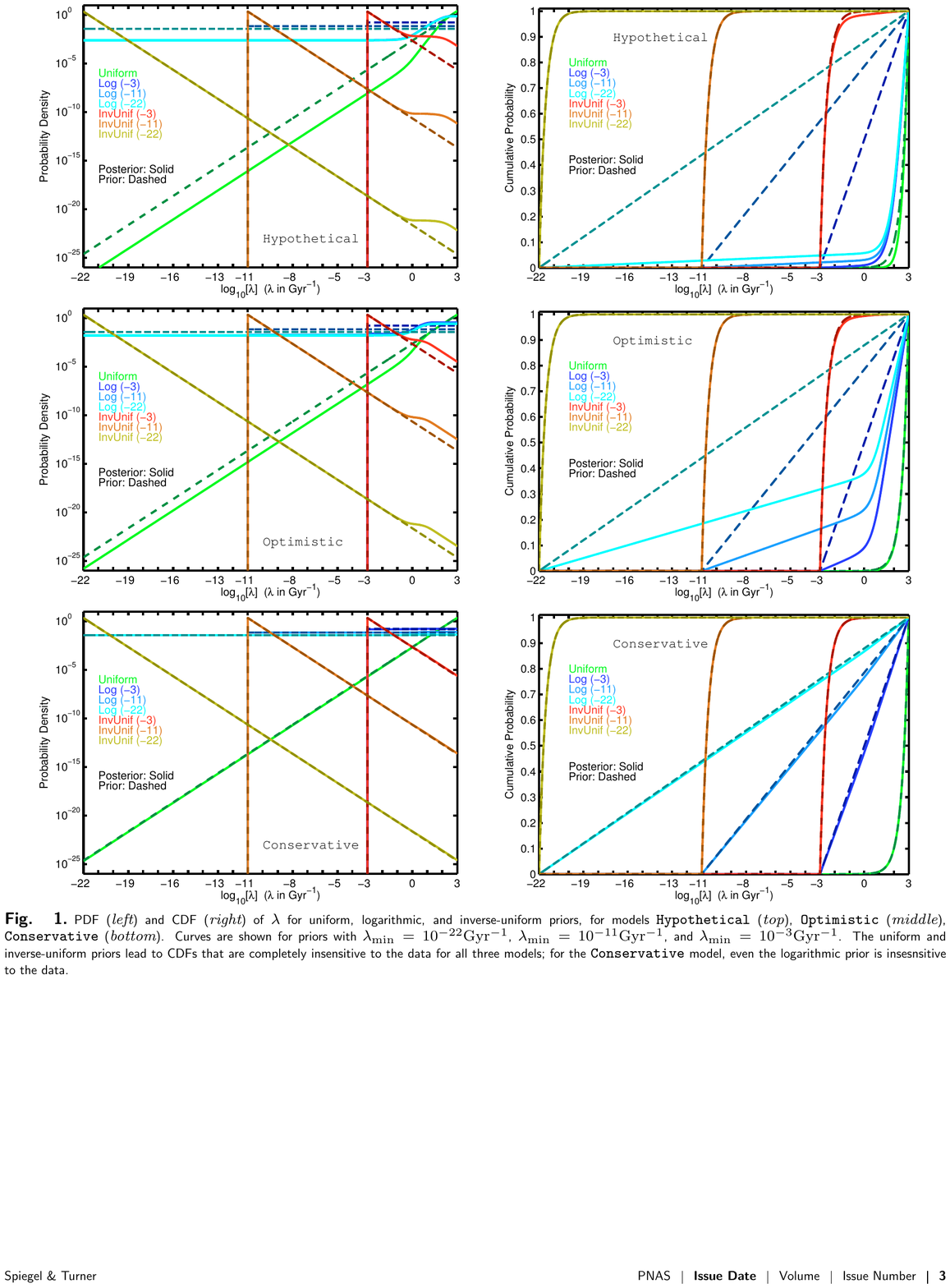}
\caption{PDF ({\it left}) and CDF ({\it right}) of $\lambda$ for
  uniform, logarithmic, and inverse-uniform priors, for models {\tt
    Hypothetical} ({\it top}), {\tt Optimistic} ({\it middle}), {\tt
    Conservative} ({\it bottom}).  Curves are shown for priors with
  $\lambda_{\rm min} = 10^{-22} \rm~Gyr^{-1}$, $\lambda_{\rm min} =
  10^{-11} \rm~Gyr^{-1}$, and $\lambda_{\rm min} = 10^{-3}
  \rm~Gyr^{-1}$.  The uniform and inverse-uniform priors lead to CDFs
  that are completely insensitive to the data for all three models;
  for the {\tt Conservative} model, even the logarithmic prior is
  insesnsitive to the data.}
\label{fig:pdfcdf_all}
\end{figure}

\clearpage

\begin{figure}
\plotone
{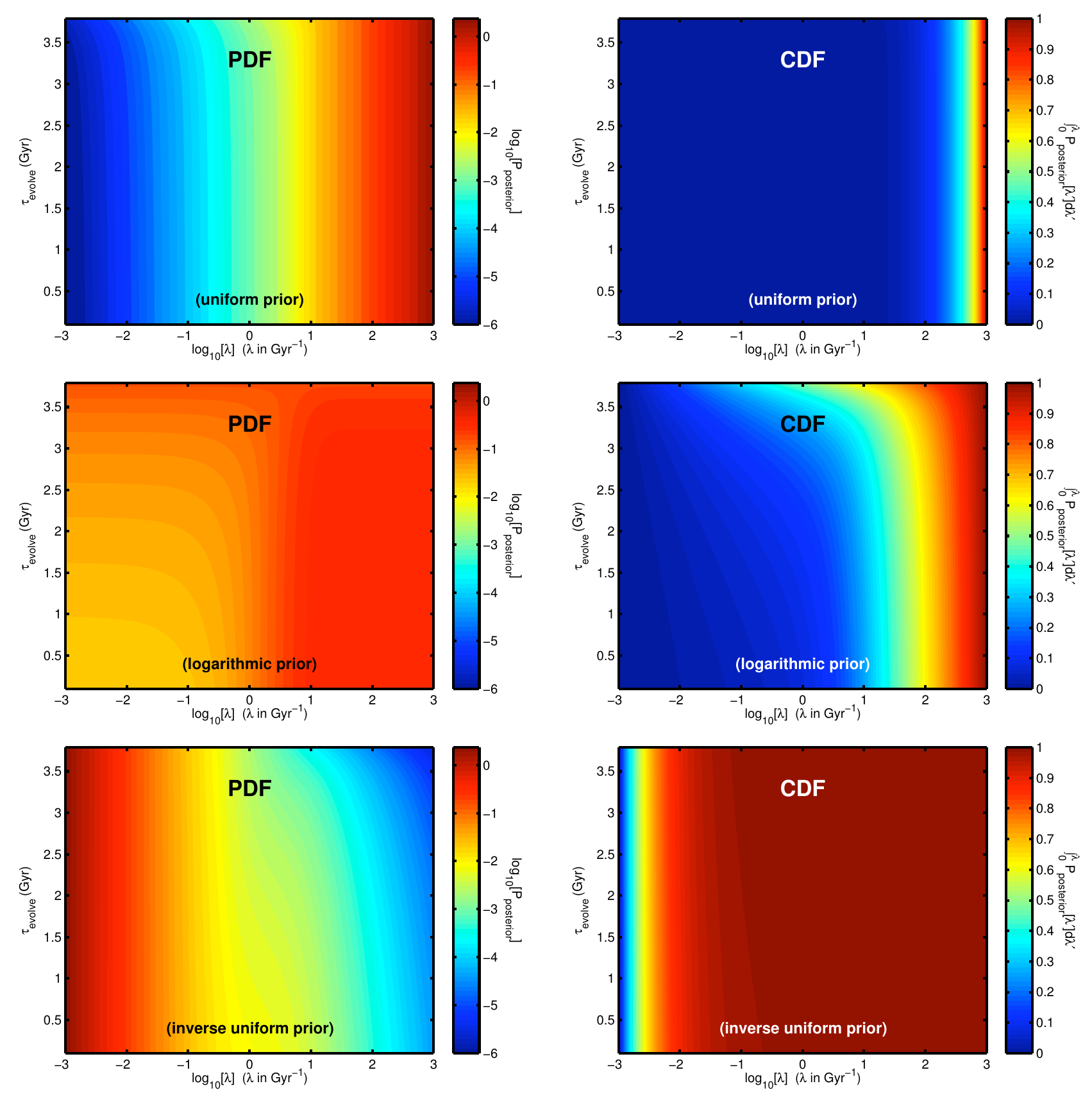}
\caption{The influence of $\delta t_{\rm evolve}$.  Uniform prior ({\it
    top}), logarithmic prior ({\it middle}), and inverse-uniform prior
  ({\it bottom}).  PDF ({\it left}) and CDF ({\it right}).  Aside from
  $\delta t_{\rm evolve}$, parameters are set to the {\tt Optimistic}
  model with $\lambda_{\rm min} = 10^{-3} \rm~Gyr^{-1}$ in the
  logarithmic and inverse-uniform cases.}
\label{fig:tau_evolve}
\end{figure}

\clearpage

\end{onecolumn}

\end{article}

\end{document}